# Integrated multi-operand optical neurons for scalable and hardware-efficient deep learning

Chenghao Feng[1,2], Jiaqi Gu[2], Hanqing Zhu[2], Rongxing Tang[1], Shupeng Ning[1], May Hlaing[2], Jason Midkiff[2], Sourabh Jain[1], David Z. Pan [2,*], Ray T. Chen[2,3,*]

[1]*Microelectronics Research Center, The University of Texas at Austin, Austin, Texas 78758, USA.*
[2]*Department of Electrical and Computer Engineering, The University of Texas at Austin, Austin, Texas 78705, USA.*
[3]*Omega Optics, Inc., 8500 Shoal Creek Blvd., Bldg. 4, Suite 200, Austin, TX 78757, USA.*
*\*chenrt@austin.utexas.edu*

**Abstract:** The optical neural network (ONN) is a promising hardware platform for next-generation neuromorphic computing due to its high parallelism, low latency, and low energy consumption. However, previous integrated photonic tensor cores (PTCs) consume numerous single-operand optical modulators for signal and weight encoding, leading to large area costs and high propagation loss to implement large tensor operations. This work proposes a scalable and efficient optical dot-product engine based on customized multi-operand photonic devices, namely multi-operand optical neurons (MOON). We experimentally demonstrate the utility of a MOON using a multi-operand-Mach-Zehnder-interferometer (MOMZI) in image recognition tasks. Specifically, our MOMZI-based ONN achieves a measured accuracy of 85.89% in the street view house number (SVHN) recognition dataset with 4-bit voltage control precision. Furthermore, our performance analysis reveals that a $128 \times 128$ MOMZI-based PTCs outperform their counterparts based on single-operand MZIs by one to two order-of-magnitudes in propagation loss, optical delay, and total device footprint, with comparable matrix expressivity.

## 1. Introduction

The optical neural network (ONN) is an emerging analog artificial intelligence (AI) accelerator that leverages properties of photons, including low latency, wide bandwidth, and high parallelism [1–3], to address the growing demand for computing power required to implement deep neural network (DNN) models. Once weight parameters are set, photonic integrated circuits (PICs) can perform tensor operations with near-zero energy consumption at the speed of light [4,5], making them an ideal platform for accelerating multiply-accumulate (MAC) operations [6]. However, the potential massive parallelism and ultra-high computing speed of ONNs are not fully unleashed with small-size photonic tensor cores (PTCs). To maximize the performance benefit of photonic computing in DNN acceleration, scalable and efficient photonic tensor core designs are in high demand.

The scalability of previous photonic tensor core designs is bottlenecked by the large spatial footprint and insertion loss [7]. For instance, an MZI-based coherent PTC [8] requires $O(m^2 + n^2)$ single-operand MZI modulators to construct an $m \times n$ matrix, consuming huge area cost to implement large tensor operations (e.g., $128 \times 128$). Moreover, the large number (~$2n$) of cascaded optical devices in the critical path of the circuit leads to unacceptable insertion loss. Even with low-loss MZIs such as thermo-optic MZIs (0.5-1 dB) [9], cascading 128 such devices will result in 64 to 128 dB propagation loss. In addition, single-operand-device-based PTCs suffer from nontrivial dynamic energy consumption to reconfigure weight parameters. Given the limited chip area and link budget, we have to serialize the matrix multiplication by repeatedly reusing small-size photonic tensor cores, which incurs much longer latency to implement one matrix-vector multiplication, potentially negating the speed advantage of ONNs over electronic analog AI accelerators [10].

Both circuit- and device-level optimizations have been explored to enhance the scalability of ONNs. Circuit-level approaches, such as the butterfly-style circuit mesh [11], have been explored to reduce hardware usage [12,13]. Moreover, compact device-level photonic tensor cores, such as star couplers and metasurfaces [5,6], have been proposed to significantly reduce the device footprint and improve the hardware efficiency of tensor operations. However, one major challenge with compact photonic circuit mesh or passive device-level tensor cores is their limited matrix representability, which usually results in accuracy degradation when implementing complicated AI tasks. To address this challenge, we suggest using active device-level photonic tensor cores, which offer the potential to achieve both high representability and high hardware efficiency. Recently there has been a trend to use multi-operand devices for vector operations, which shows great potential to achieve efficiency and scalable breakthroughs. We partition the phase shifter into multiple small segments, each being independently controlled. By leveraging the underlying device transfer function, we can then realize vector operations with nearly the same device footprint and tuning range as the single-operand one. In this work, for the first time, we officially name this photonic structure a multi-operand optical neuron (MOON). Prior work has proposed a microring-based MOON and showed its advantages over standard single-operand micro-ring in neuromorphic computing through simulation [14]. In this work, we introduce a new broadband device in this MOON-family, a multi-operand MZI (MOMZI), and experimentally demonstrate its superior efficiency and scalability for next-generation photonic neuromorphic computing.

In this work, we customize a MOMZI, whose modulation arm is controlled by multiple independent signals, and leverage its transmission to realize vector-vector dot-product. A $k$-operand ($k$-op) MOMZI can be used as a length-$k$ vector dot-product engine, directly saving the device usage by a factor of $k$ compared to single-operand MZI arrays [8]. Note that the device footprint and tuning range keep constant and will not scale with $k$. By combing the result from multiple $k$-op MOMZIs, we can efficiently scale up to operations with a large vector length with near-constant insertion loss. Using devices from foundry process design kits (PDKs) [15], 128×128 photonic tensor cores based on our MOMZIs show a 6.2× smaller total device footprint, 49× lower optical delay, and >256 dB lower propagation loss than previous single-operand MZI arrays [8]. We experimentally demonstrated the representability and trainability of an ONN constructed by 4-op MOMZIs on the street view house number (SVHN) recognition task [16], achieving a measured accuracy of 85.89% with 4-bit voltage control precision. Our proposed MOMZI-based photonic tensor core enables the implementation of high-performance and energy-efficient neuromorphic computing with a small device footprint, low propagation delay, and low energy consumption.

## 2. Multi-operand optical neurons

A typical photonic tensor core to implement MAC operation is in Fig. 1(a), which contains photonic components to generate input signals, the weight matrix, and the outputs. $n$ high-speed modulators are needed in an $n$-input, $m$–output layer. Depending on the weight mapping approach, one needs $\frac{m(m-1)+n(n-1)}{2} + \max(m,n)$ [8] or $m \times n$ active photonic components [17] to implement a $m \times n$ weight matrix. Furthermore, ~$2n$ active devices are cascaded in one optical path, resulting in nonnegligible propagation loss and requiring more laser power to drive the photonic neural chip.

In this study, we propose a novel approach to reduce the optical component usage by implementing the multiply-accumulate (MAC) operation using an array of multi-operand-modulator-based optical neurons (MOONs), as shown in Fig. 1(b). Depending on the area and reliability concerns, one MOON can be a multi-operand active photonic device of any waveguide structure, such as MZI modulators and microring modulators. As illustrated in Fig. 1(c), each row of the layer is divided into $n^* = \frac{n}{k}$ $k$-operand modulators, and the output of each $k$-operand modulator is accumulated using on-chip combiners or multiplexers to compute the final output of each row. Consequently, the total number of MOONs required for an $n$-input, $m$-output layer is $\frac{mn}{k}$, significantly reducing the number of active optical components.

Unlike conventional PTCs designed for GEMM, the nonlinearity transfer function between the electrical signal and the transmission of the MOON needs to be considered when training DNN models. The input vector $x_{in}$ is encoded as the amplitude of the optical signals and will also be partitioned into $n^*$ segments $x_{in} = (x_{in}^1, x_{in}^2, \ldots, x_{in}^n)$. Thus, the output signals of one layer can be expressed as follows:

$$x'_{out} = f(Wx_{in}) = \begin{pmatrix} \Sigma_{i=1}^{p=n^*} f\left(\Sigma_{j=1}^{k} g\left(W_{1,j+(i-1)k}, x_{in}^{j+(i-1)k}\right)\right) \\ \Sigma_{i=1}^{p=n^*} f\left(\Sigma_{j=1}^{k} g\left(W_{2,j+(i-1)k}, x_{in}^{j+(i-1)k}\right)\right) \\ \vdots \\ \Sigma_{i=1}^{p=n^*} f\left(\Sigma_{j=1}^{k} g\left(W_{m,j+(i-1)k}, x_{in}^{j+(i-1)k}\right)\right) \end{pmatrix}. \qquad \text{Eq. (1)}$$

where function $f(\cdot)$ represents the relationship between the total phase shift or amplitude response of all the operands, whereas $g(w_i, x_i)$ is determined by the weight/signal encoding way and each operand's phase/amplitude response. For simplicity, we can use $V(w_i, x_i) = w_i \cdot x_i$ directly. As depicted in Fig. 1(e)-(g), $w_i$ can be encoded by programmable resistances (e.g., memristors or phase change materials [18]), tunable electrical amplifiers/attenuators, or the length of modulation arms if the weights are fixed. $x_i$ refers to the input current or voltage signals from input sources or the previous layer. After obtaining the transfer function of MOON (Eq. 1), one can deploy them in commercial deep learning platforms, e.g., Pytorch, to train MOON-based PTCs.

Our MOON-based PTC significantly improves computational efficiency compared to previous GEMM-based PTCs [8]. For example, a $k$-operand MOMZI has a similar device footprint and dynamic tuning range to a single-operand MZI, but it can implement $k$ MACs. This outperforms a single-operand MZI in area- and energy- efficiency since it can only perform approximately one MAC operation per device in single-operand MZI-based PTCs. The advantage of MOONs lies in their ability to perform multiple MAC operations using a single device, making them more computationally efficient than previous ONNs.

Moreover, as shown in Fig. 1(c), only one MOON is cascaded in one optical path of our circuit architecture, resulting in much smaller propagation loss compared to MZI-based or microring-based ONNs, where $2n + 1$ MZIs or $n$ microrings are cascaded. As a result, we can deploy compact but lossy optical modulators, e.g., plasmonic-on-silicon modulators [19], as MOONs in our PTC, trading higher insertion loss for a much smaller chip footprint and lower modulation power. Detailed performance evaluations will be provided in our discussions.

## 3. Multi-operand-MZI-based optical neural network

In this work, we demonstrate the use of $k$-operand MZI modulators as the fundamental building blocks for constructing our MOMZI-PTC. Figure. 1(d) shows the structure of a MOMZI. Unlike the traditional MZI modulators with one or two phase modulators, a $k$–op MOMZI has $k$ active phase shifters on each modulation arm, and each phase shifter is controlled by an independent signal. This structure is similar to lumped-segment MZIs used in optical communications [20], but the driving signals on each operand are independent and analog. For

MZI modulators with dual modulation arms, the total number of operands can increase to $2k$ to enable both positive and negative phase shifts. Suppose each shifter contributes to a phase shift $\phi_i$, the output intensity of a MOMZI can be expressed as:

$$y_i = f(\Sigma \phi_i) = f\left(\Sigma_{i=0}^{k}\phi_i^+ - \Sigma_{i=0}^{k}\phi_i^-\right) = \cos^2\left(\frac{\Sigma_{i=0}^{k}\phi_i^+ - \Sigma_{i=0}^{k}\phi_i^- + \phi_b}{2}\right) \quad \text{Eq. (2)}$$

where $f(\cdot) = \cos^2(\frac{\cdot}{2})$, $\phi_i^+$ denotes the $i$-th phase shifters on the upper arm of the modulator, and $\phi_i^-$ denotes that on the lower arm. Consequently, positive weight signals are encoded on upper modulation arms, while negative weight signals are encoded on the lower ones. $\phi_b$ is the phase bias when no input signals are operated on the modulation arms, which is used to tune the transfer function of the MOMZI.

The modulation mechanism of the MOMZI plays a critical role in determining their transfer function with an input voltage signal. As shown in Fig. 2, the transfer function of MZI modulators using the same foundry [15] can exhibit sinusoidal, quadratic (linear field intensity response), or other nonlinear transfer functions with the operating voltage $V$. The specific shape of the transfer function depends on the modulation mechanism ($\phi_i^{\pm}(V)$) and the modulator's waveguide structure ($f(\cdot)$). By optimizing these parameters, one can customize the transfer function of the MOMZI to realize certain nonlinear activation functions of DNNs. We will discuss this hereinafter.

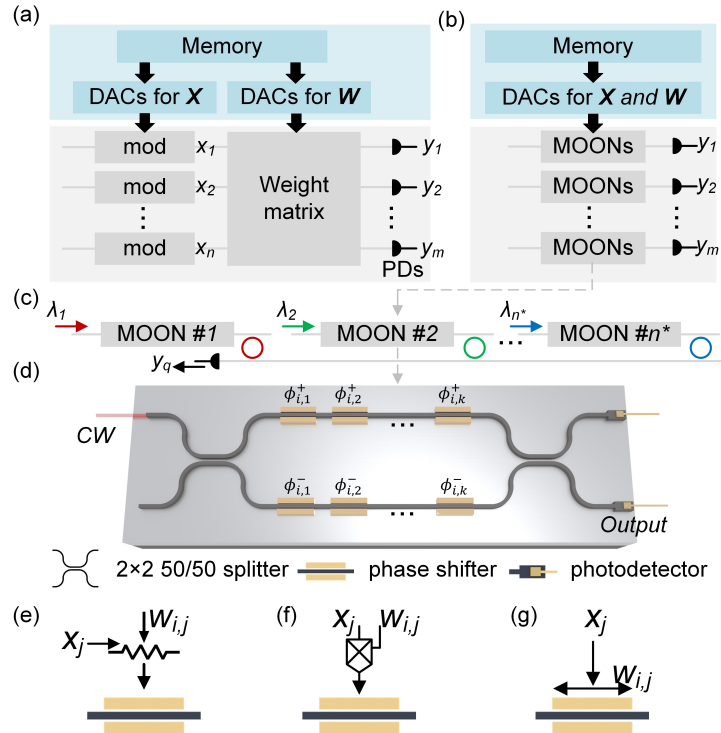

**Fig. 1 General architecture of the MOON-based photonic tensor core.** (a) A conventional photonic tensor core based on single-operand modulators, which has an array of input modulators and $O(mn)$ photonic devices to construct the weight matrix. (b) Schematic of the MOON-based PTC to implement an $n$-input, $m$-output layer. (c) Schematic of MOON array to implement a $n \times 1$ vector product operation, which requires $n^* = \frac{n}{k}$ $k$-operand MOONs. The output is obtained by accumulating the output signal of each MOON using multiplexers. (d) Schematic of a $k$-operand MOMZI-based MOON, which consists of $k$ operands on each arm. There are various approaches to encoding weight signals $w_i$ and input signals $x_i$ on each modulation region. For instance, one can use (e) programmable resistances to encode $w_i$ and current signals to encode $x_i$, or (f) tunable amplifiers/attenuators to encode $w_i$ and voltage signal to encode $x_i$, or (g) adjust modulation length to encode fixed weight signals and voltage signals to encode $x_i$.

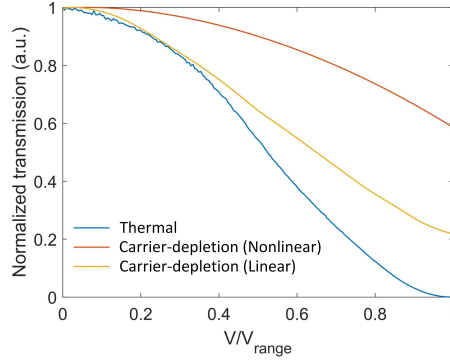

**Fig. 2 Transfer function of different MZI modulators under different modulation mechanisms.** All the data are experimental data from our measurement or the process design kit (PDK) model [15] on Lumerical Interconnect. $V_{range}$ is the maximum allowed operating voltage.

Supposing the dot product information $w_i \cdot x_i$ is directly encoded as the operating voltage $V_i$ on each operand of the MOMZI, we can rewrite Eq. (2) as Eq. (3):

$$x'_{out} = f(Wx_{in}) = \frac{1}{2}\begin{pmatrix} \Sigma_{i=1}^{n^*}\left(\cos\left(\Sigma_{j=1}^{k}\phi(W_{1,j+(i-1)k}\, x_{in}^{j+(i-1)k}) + \phi_b^{1,i}\right)\right) \\ \Sigma_{i=1}^{n^*}\left(\cos\left(\Sigma_{j=1}^{k}\phi(W_{2,j+(i-1)k}\, x_{in}^{j+(i-1)k}) + \phi_b^{2,i}\right)\right) \\ \vdots \\ \Sigma_{i=1}^{n^*}\left(\cos\left(\Sigma_{j=1}^{k}\phi(W_{m,j+(i-1)k}\, x_{in}^{j+(i-1)k}) + \phi_b^{m,i}\right)\right) \end{pmatrix} + b \quad \text{Eq. (3)}$$

In Eq. (3), positive or negative phase shifts are achieved by applying the operating voltages to each phase shifter's upper or lower arm. The phase bias $\phi_b^{p,i}$ of the $i$th MOMZI on row $p$ of MOMZI-PTC can be adjusted to improve the expressivity of our neural architecture. The constant $b = \frac{n}{k}$ can be eliminated after photodetection. Using Eq. (3). we can model the MOMZI on commercial deep learning platforms, e.g., PyTorch, making it practical to train and deploy the DNN.

## 4. Experimental results

In this study, we designed and fabricated a 4-op MOMZI that is capable of implementing a 4×1 vector operation on the silicon photonics platform. The chip layout was drawn and verified using Synopsys OptoDesigner (version 2021) and then fabricated by AIM Photonics. The schematic of the MOMZI is illustrated in Fig. 3(b), while Fig. 3(a) shows close-up images of its components, including phase shifters, 50-50 directional couplers, and photodetectors.

We use two phase shifters on each modulation arm to enable both positive and negative weights during training. In experiments, we encode $\phi_i \propto |w_i^{\pm}| \cdot x_i$, where $w_1^+$ and $w_2^+$ are positive and $w_1^-$ and $w_2^-$ are negative, $w_1$ and $w_2$ are encoded on the upper arm, while $w_1^-$ and $w_2^-$ are encoded on the downer arm. The transfer function of our modulator can then be written as

$$T = f(\Sigma_i \phi_i + \phi_b) = \cos^2(\phi_1(w_1^+ \cdot x_1) + \phi_2(w_2^+ \cdot x_2) - \phi_3(w_1^- \cdot x_3) - \phi_4(w_2^- \cdot x_4) + \phi_b) \quad \text{Eq. (4)}$$

We tune one additional phase shifter on the upper arm to let $\phi_b \approx \frac{\pi}{2}$ to obtain a relatively linear and balanced output range.

The schematic of the testing setup is illustrated in Fig. 4. Continuous-wave (CW) light is coupled to the chip through an edge coupler. The MOMZI's phase shifters are programmed using a high-precision multi-channel digital-to-analog converter (DAC). The on-chip photodetector, along with an off-chip transimpedance amplifier (TIA), converts the output optical signal to electrical voltage outputs. These converted electrical outputs will subsequently be read using oscilloscopes. A microcontroller is used to program the electrical signals that represent $w_i \cdot x_i$ to the DAC and read the output signals in this work. We use computers to process the measurement data, train the DNN parameters, and implement the DNN model. Notably, current fabrication and co-packaging technologies enable the integration of electrical control circuits and the laser on a single substrate [21] or a single chip [22], resulting in much higher compactness, shorter interconnect paths, and higher efficiency.

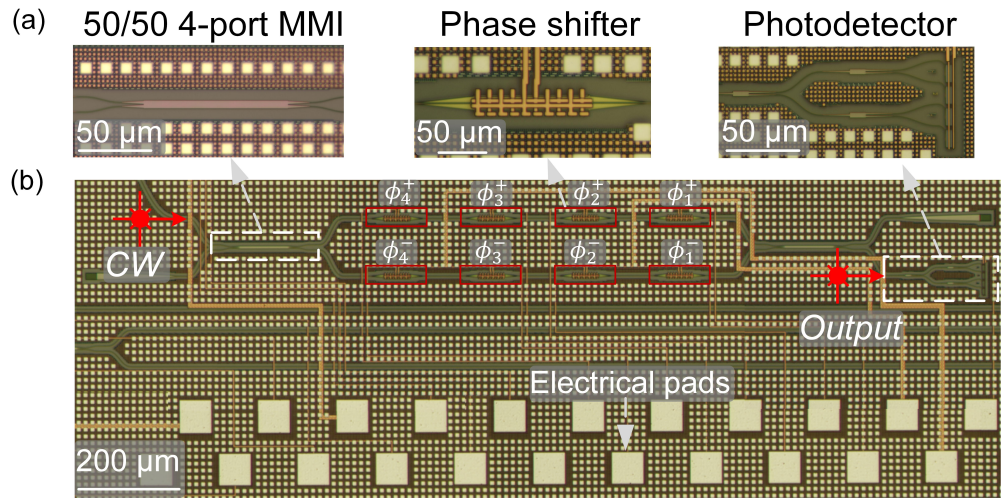

**Fig. 3 Schematic of the 4-operand MOMZI.** The micrographs of necessary optical components are highlighted in (a) and the full schematic of the MOON is shown in (b).

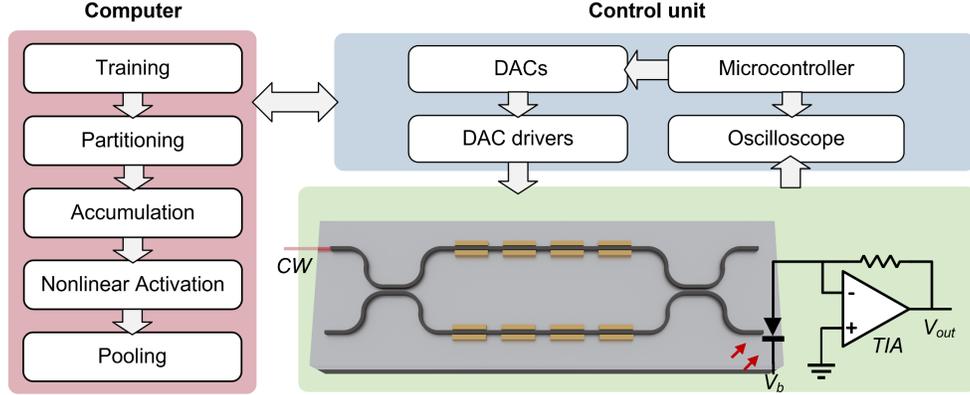

**Fig. 4 Experimental setup of MOMZI-ONN.** (a) Schematic of our MOMZI-ONN test flow. The entire tensor operation is first partitioned into multiple 4×1 blocks, and each block is implemented optically on a 4-op MOMZI. (b) The weight parameters and the input signals are programmed by a multi-channel digital-to-analog converter (DAC). Shown in (c), the output optical signals are converted to photocurrents using on-chip photodetectors. We use an off-chip TIA to convert the output photocurrent to electrical signals, which are then read by the oscilloscope. Both the oscilloscope and the DAC are controlled by a microcontroller. The tensor operation results are provided to the computer for data processing in order to train and deploy the DNN.

In this work, we construct a CNN with our MOMZI and benchmark its performance on a street view house number (SVHN) dataset. It is more complicated than the MNIST dataset [16] since each image contains color information and various natural backgrounds. To perform convolutional operations with our PTCs, we employ the widely-used tensor unrolling method (im2col) [23]. Large-size tensor operations are partitioned into 4×1 blocks and mapped onto our MOMZI. We first calibrate the behavior of each phase shifter for training and model it using Eq. 4, as shown in Figure 5(a). Based on the chip measurement data, our proposed hardware-aware training framework can efficiently train the ONN weights while being fully aware of all the physical non-idealities during optimization, e.g., process variations, thermal crosstalk, and signal quantization [14]. The dynamic noises are also measured (shown in Fig. 5(b)) and added to the training framework to improve the robustness of ONNs. Then we map the trained weights to our MOMZI. Finally, we evaluate the task performance of our photonic neural chip on different ML tasks, where partial accumulation, nonlinearity, and other post-processing operations are offloaded to the digital computer. Figure 5(c) illustrates the network structure for training our MOMZI-ONN.

Our experiments show that under 4-bit voltage control resolution (16 phase shift levels for each operand), the inference accuracy of the CNN reaches ~85.89% in our experimental demonstration. The confusion matrix depicting the prediction results is shown in Fig. 5(d). Figure 5(e)(f) shows the tested probability distribution of different street-view numbers. As a reference, we can achieve 90.6% accuracy using an ideal CNN model with the same network structure on 64-bit computers. One can improve the task performance of MOMZIs using operands with more linear phase responses and higher control precision, which will be shown hereinafter.

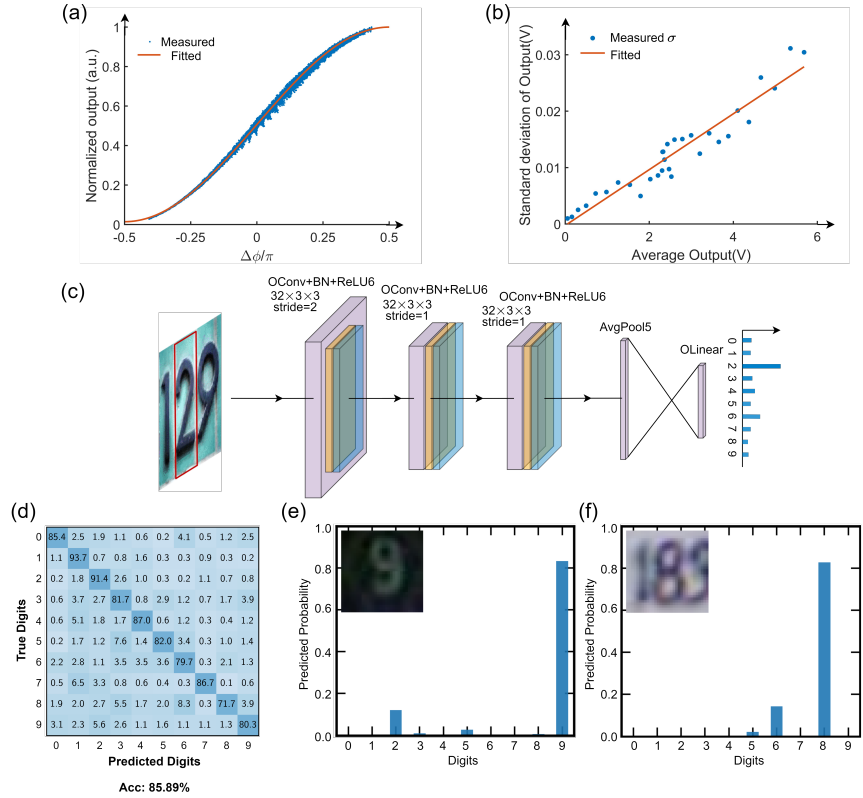

**Fig. 5 Experimental result of street view house number (SVHN) recognition with the MOMZI-ONN.** (a) Our measured output data and curve fitting for training the MOMZI. The tuning range of the total phase shift of four operands is $[-\frac{\pi}{2}, \frac{\pi}{2}]$. (b) Dynamic noise analysis of output signal of MOMZI, the measured standard deviation of the dynamic noise is ~0.5%. (c) Structure of the CNN, the convolution is realized by MOMZIs with im2col approach. The first convolutional layer has one input channel and 32 output channels with a stride of 2. The subsequent two convolutional layers have 32 input/output channels with a stride of 1. After adaptive average pooling, we use a linear classifier with 10 outputs for final recognition. (d) Our measured output data and curve fitting for training the MOMZI-ONN. The tuning range of the total phase shift of four operands is $[-\frac{\pi}{2}, \frac{\pi}{2}]$. (c) The confusion matrix of the trained MOMZI-ONN on the SVHN dataset shows a measured accuracy of 85.89%. (e) and (f) show the predicted probability distribution of our MOMZI-ONN on two selected test digits in the SVHN dataset.

## 5. Discussion

**Expressivity evaluation.** Our MOMZI-ONN exhibits comparable trainability and expressivity with ONNs designed for GEMMs with $k$ times fewer optical component usage ($k$ is the number of operands). By explicitly modeling the transfer function of the MOMZI during ONN training, we can efficiently learn the mapping from the software model to the MZI devices. Here, we simulate the task performance of our MOMZI-ONN with different numbers of operands on the SVHN dataset using the same NN model and control precision. An ideal CNN model with the same model architecture is also trained as a reference. In the evaluation, the phase response of each operand is $\phi = \gamma V$, which can be realized on linear phase shifters such as lithium niobate EO phase shifters [24]. In simulations, we add a phase bias $\phi_b = \frac{\pi}{2}$ to enable a balanced output range. The evaluation results are shown in Fig. 6, showing that our MOMZI-ONNs can achieve >91% accuracy on the SVHN dataset, which has a <0.6% accuracy difference compared to the ideal CNN model. It should be noted that the task performance of MOMZI-ONN is insensitive to the number of operands once we properly normalize the operands. Moreover, the number of active photonic devices to implement an $n$-input, $m$-output linear layer are $\frac{mn}{k}$ $k$-operand MOMZIs. Therefore, ONNs based on MOMZIs with a large number of operands will significantly reduce the hardware cost without accuracy loss.

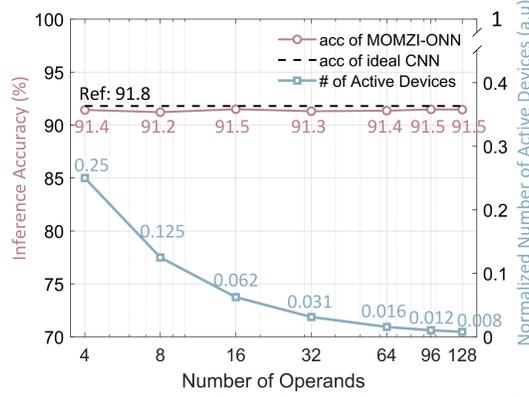

**Fig. 6 Task performance and hardware cost of MOMZI-ONN on SVHN dataset.** Inference accuracies of MOMZI-ONNs with different operand numbers are shown. Using the same neural network structure, the accuracy of an ideal CNN model is 91.8%. The normalized total number of active photonic devices with different operand numbers of MOMZI-ONNs compared to ideal CNN models are shown. We suppose the matrix size is $n \times n$, the number of microring- and MZI-ONNs are normalized to 1.

**Propagation delay and loss.** By minimizing the number of cascaded optical components in the critical path of PTCs, MOMZI-PTC outperforms single-operand MZI-PTC in both propagation loss and optical delay by one to two orders of magnitude. In this work, we evaluate the propagation delay and loss of MOMZI-PTC using the foundry's process-design-kit (PDK) libraries. The parameters of optical devices are given in Table. S1. As shown in Fig. 1(c), the MOMZIs in one optical path are placed parallelly in our PTC, so the insertion loss and propagation loss contributed by lossy MZIs will not accumulate when the size of the DNN model increases. As a result, the optical delay and the propagation loss of a MOMZI-PTC with n-inputs and m- outputs can be calculated as follows:

$$\tau_{MOMZI-PTC} = \frac{n_g}{c}(L_{MOMZI} + L_{combiner}) \quad \text{Eq. (5)}$$

$$IL_{MOMZI-PTC} = IL_{MOMZI} + IL_{combiner} \quad \text{Eq. (6)}$$

In Eq. (5), $n_g = 4.3$ is the group index of silicon waveguides. $L_{MOMZI}$ is the length of the MOMZI, which depends on the operands and the waveguides used to connect these operands. Since the tuning ranges of a MOMZI and a single-operand MZI are the same, the total length of the operands of MOMZI should also be the same as the length of a high-speed electro-optic (EO) modulator. Here we assume the distance between each operand to be $d = 10 \ \mu m$ based on the device layout of a recently-published two-operand 10- $\mu m$ -radius microring modulator [25]. $L_{combiner}$ is the length of the on-chip combiners/multiplexers, for microring-filter-based multiplexers, $L_{combiner} = \frac{n}{k} L_{ring}$. $IL_{MOON}$ is the insertion loss of one multi-operand modulator, $IL_{combiner}$ is the total insertion loss of the combiner, which is $\frac{n}{k} IL_{ring}$ with add-drop microrings as multiplexers. Increasing the operand number k can potentially reduce both the IL and propagation delay.

On the other hand, the propagation loss of single-operand MZI-PTC can be estimated as $(n + m + 1)IL_{MZI(ls)} + IL_{MZI(hs)}$, while the total device length can be expressed as

$(n + m + 1)L_{MZI(ls)} + L_{MZI(hs)}$. Here, $MZI(hs)$ denotes the high-speed EO modulators for input signal encoding, and $MZI(ls)$ is the TO switch for weight encoding. Because the tuning range and the modulation mechanism of the MOMZI should be the same as that of the input EO modulator, we let $IL_{MOMZI} = IL_{MZI(hs)}$. The model parameters are available in Table. S1.

The results presented in Fig. 7(a) and 7(b) demonstrate that using the same component library [15], our MOMZI-PTC can achieve an optical delay that is approximately 49 times lower than that of a single-operand MZI-PTC. Furthermore, the propagation loss of our MOMZI-PTC is ~257 dB lower than that of the single-operand MZI-PTC, which results in lower laser power requirements to drive the ONN and a lower response time.

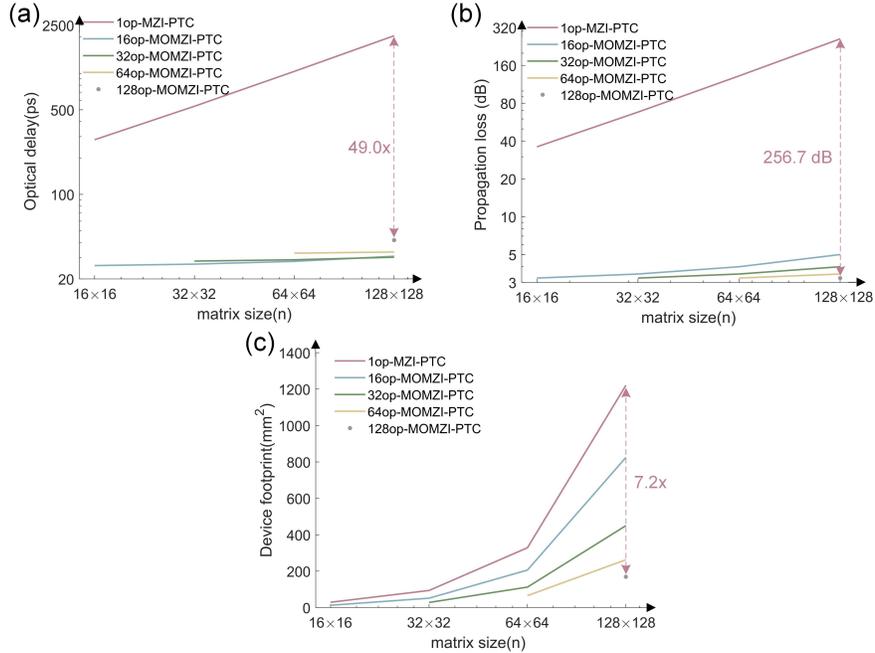

**Fig. 7 Performance analysis of MOMZI-PTC and comparison with single-operand (1op) MZI-PTC** [8] **using foundry PDKs** [15]**.** MOMZIs with different operand numbers are shown. Here we suppose the circuit structure of the MZI-based PTC is Clement-style [26]. (a) Optical propagation delay in log scale. (b) Optical propagation loss. (c) Device footprint.

**Footprint.** Our MOMZI-PTC significantly improves the area efficiency and reduces the number of active devices compared to a single-operand MZI-PTC [8]. Unlike single-operand active devices such as single-operand MZI, our $k$-op MOMZI is capable of implementing $k$ dot products and $k − 1$ additions, which results in a much higher hardware efficiency in terms of #MAC/device. The total device footprint of $k$-op MOMZI-PTCs can be estimated using Eq. (7):

$$S_{MOMZI-PTC} = \frac{m \times n}{k} S_{MOMZI} + S_{combiner} \qquad \text{Eq. (7)}$$

where we assume a distance of $d = 10 \ \mu m$ between neighboring operands. Suppose the device footprint of a high-speed MZI modulator is $S_{MZI(hs)} = L_{MZI(hs)} \cdot W_{MZI(hs)}$. The device footprint of one $k - op$ MOMZI can then be estimated as $S_{MOMZI} = (L_{MZI(hs)} + (k-1)d) \cdot W_{MZI(hs)}$. Figure. 7(c) shows the estimated device footprint of MOMZI-based PTC and single-operand MZI-PTC based on our assumptions. The estimated device footprint of MOMZI-PTC and MZI-PTC is shown in Fig. 7(c). When the matrix size is $128 \times 128$, our 128-op MOMZI-PTC consumes ~127× fewer MZI modulators, leading to ~6.2× footprint reduction compared to single-operand MZI-PTC [8] with the same matrix size and optical component selection.

From Eq. (7) and Fig. 7(c), MOMZI-based PTC will be more area efficient with a larger number of operands $k$ on each MOMZI. Previous has shown that a 10-μm-radius silicon-based microring modulator can be divided into 32 independent active segments using a 45-nm technology node [27], where each operand only consumes 2μm in length. This means that an MZI-modulator with a 1.6 mm-length modulation arm can support up to 800 operands using current layout technology, which should be comparable with other analog electronic tensor cores in scalability, e.g., $256 \times 256$ memristor-based crossbar arrays [28].

Another big advantage of the proposed MOON-based PTC is its superior compatibility with compact, high-speed optical modulators, even with high insertion loss, e.g., plasmonic-on-silicon modulators with only 15 μm modulation length and 11.2 dB $IL$ [5]. The fundamental reason is the small number of cascaded devices in the critical path. Figure 7 shows the normalized device footprint compared with silicon-based MZI-PTCs, which shows the plasmonic-on-silicon-MOMZI-PTC can reduce the footprint by 177× compared to single-operand silicon-MZI-based PTC. Single-operand MZI-PTCs are not compatible with these compact high-loss modulators because there are $2n + 1$ MZIs in the critical path. Using compact high-loss modulators for weight configuration will lead to significant propagation loss and require high laser power to drive the neural chip.

Finally, the hardware cost of MOMZI-PTC can be further optimized with operand pruning strategies. To implement an FC layer in DNN models, especially with sparse matrices [29], we only need to encode non-zero weights on the MOMZIs. The operands of MOMZIs with zero weight values can be either removed from the device to save footprint or power-gated to reduce energy consumption. Sparsity-aware training [30] can be applied to prune redundant MOMZI operands while maintaining task accuracy.

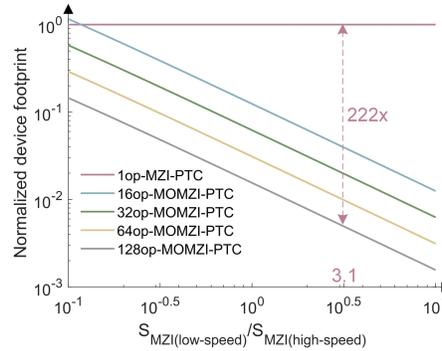

**Fig. 8 Normalized footprint of MOMZI-based PTC using scaling technologies.** The x-axis is the ratio between the area of low-speed silicon-based TO MZI ($550 \times 125\ \mu m^2$) and high-speed MZI. Using compact plasmonic-on-silicon high-speed modulators (~$220 \times 100\ \mu m^2$) [19], $S_{MZI(low-speed)}/S_{MZI(high-speed)} \cong 3.12$, and a 128-op MOMZI-PTC consumes a 222x smaller footprint than single-operand MZI-PTCs using silicon-based MZI modulators. For simplicity, we assume the entire waveguide length for connecting the operands is the same as the total length of the operands of MOMZIs, so $S_{MOMZI} = 2S_{MZI(high-speed)}$.

**Energy efficiency.** MOMZI-based PTC is a more energy-efficient alternative to single-operand MZI-PTCs for implementing large-tensor-size operations due to its lower propagation loss, which allows it to consume over 256 dB less laser power. The total power consumption of MOMZI-PTC for computing comprises the power required to drive the lasers, modulators, and photodetectors and for biasing the MOMZI, as well as the power needed to drive the digital-to-analog converters (DACs) and analog-to-digital converters (ADCs). The silicon-based carrier-depletion MZI's modulation energy consumption in previous work can achieve ~146 fJ/bit [31]. Furthermore, the power to bias the MOMZI is ~2.5 mW per phase shifter if we use thermal phase shifters from foundry PDKs [32].

Using the parameters of existing technology provided in Table. S3, the optical part of MOMZI-PTC, accounts for <9% of total power consumption when the tensor size is 128 × 128. The power breakdown analysis in Fig. 8 indicates that our 128-op MOMZI-PTC can achieve ~56 TOPS/W at a 10 GHz clock rate, 100% higher than existing analog electronic tensor cores [33] with 100x faster operating speed. Currently, the energy efficiency of MOMZI-PTC is dominated by data converters such as ADCs. This work employs an 8-bit, 10 GSPS ADC that consumes 39 mW per channel [34].

The energy efficiency can be further improved to ~ 604 TOPS/W using emerging high-speed and energy-efficient data converters and EO modulators. Recent advances in energy-efficient active optical components, such as the plasmonic-on-silicon modulator that consumes approximately 0.1 fJ per bit modulation energy at 50 GHz operating frequency, have made it possible to reduce the power consumption of MOMZI further [19]. The power to bias the MOMZI can be decreased to zero with phase change materials or nano-opto-electro-mechanical devices [35,36]. Using energy-efficient modulators, the energy consumption of the optical computing part only accounts for <3% of the total power consumption, showing that large-size MOMZI-PTC will not bring scalability issues due to excessive laser power. Moreover, we can use energy-efficient analog content-addressable memory (ACM) to replace the ADCs [37], reducing the power consumption of ADCs by ~33x. The final power breakdown of MOMZI-PTC for computing shows our MOMZI-PTC can achieve a competitive energy efficiency of ~604 TOPS/W, 20x higher than existing memristor-based analog electronic tensor cores [33]. More details of our power analysis are provided in supporting information note 2.

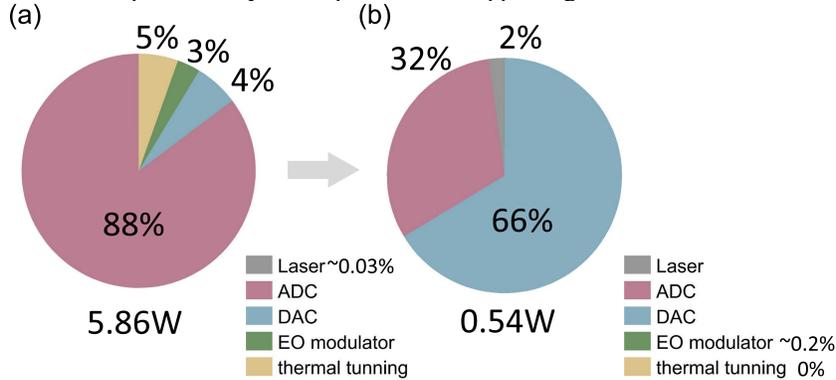

**Fig. 9 Power breakdown of a 128 × 128 photonic tensor core implemented by 128 128-op MOMZIs using existing technology(a) and emerging technology(b).** (a) The total power of the MOMZI-ONN is 5.7W at 10 GHz clock rate (56 TOPS/W). (b) Using emerging technologies, we use ADC-less designs (e.g., magnetic-tunnel-junction (MTJ)-based analog content-addressable memory (ACAM) [37,38]) to boost the energy efficiency to ~604 TOPS/W.

In addition, our $k$-op MOMZI-PTC can reduce the weight reconfiguration energy by $k$ times compared to single-operand-device-based PTCs, which will bring considerable energy efficiency improvement, especially when the photonic tensor cores need to be frequently reconfigured to map a large number of matrix blocks in DNNs. The number of active optical devices in our MOMZI-PTC is only $O(\frac{mn}{k})$, which is $k$ times fewer than that of PTCs with single-operand devices ($O(mn)$ [17] or $O(\max(m^2, n^2))$ [39]) This feature of MOMZI-PTC is essential to implement modern DNNs, where weight loading takes nontrivial hardware costs [22].

**Nonlinearity engineering.** The nonlinearity of MOONs can be customized in various dimensions to achieve a desired activation function, potentially saving power for doing activation functions electronically. The built-in nonlinearity of MOON is contributed by the weight/signal encoding way and the nonlinear transfer function of the optical modulator with the input voltage. To customize such built-in nonlinearity, one can add electrical or optical components before or after photodetection to alter the optical outputs to implement the activation function. Previous work has widely investigated this approach [41–43]. Typically, one can add saturable absorbers before photodetection with a linear optical modulator [44] to construct a ReLU-like MOON, reducing the hardware cost to realize activation functions electronically.

Depending on the transfer function of the MOON, the weight encoding approach can be designed to enable high-speed dynamic tensor operations beyond ones with stationary weights. Dynamic tensor operations mean both the inputs and the weights can be updated at high speed, which is crucial in emerging applications, such as the self-attention operation in Transformer [45] and on-chip training tasks for intelligent edge learning. A specific example of an optical modulator with a linear field response region with voltage ($|\Delta E_{out}| \propto \Delta V$) is provided here. Suppose the electrical modulation signal of the modulator is bidirectional; then, one can use two MOONs and one differential photodetector to implement high-speed vector-to-vector operations. As shown in Fig. 10, the weight and input voltage signal $w_i$ and $x_i$ are encoded with the same phases on operand $i$ of the upper modulator, and high-speed signals $w_i$ and $x_i$ with opposite phases encoded on operand $i$ of the downer modulator. After differential photodetection, one can obtain the output current signal as:

$$I_- = I_0 + \alpha(\Sigma(w_i - x_i)^2)$$
$$I_+ = I_0 + \alpha(\Sigma(w_i + x_i)^2)$$
$$I_{out} = I_+ - I_- = 2\alpha\Sigma(w_i \cdot x_i) \quad \text{Eq. (8)}$$

where $\alpha$ is the modulation efficiency of each operand. $I_0$ is the output intensity of the modulator at the biased point. Compared to MOONs that use memristors to encode stationary weights, the dual-linear-modulator-based MOON shown in Fig. 10 can enable high-speed weight reprogramming/updates to implement high-speed dynamic tensor operations. One can investigate more efficient signal encoding approaches of MOONs to support more types of tensor operations in state-of-the-art DNNs.

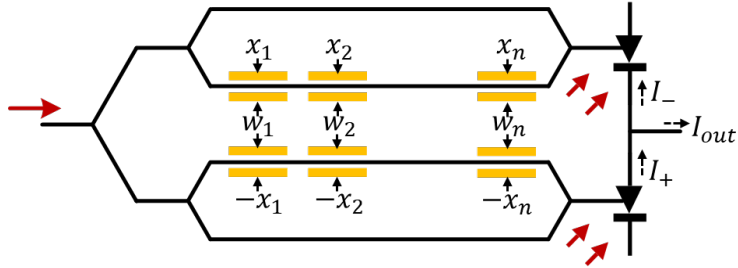

**Fig. 10 A MOON with two linear modulators for matrix-matrix multiplications.** The optical output power of each modulator is proportional to the electrical input power or $V^2$. Here we apply differential input signals $\pm x_i s$ on upper/lower modulators, and put weight signals $w_i s$ on both upper/lower modulators. The output power after differential photodetection is then proportional to $\sum_1^n w_i \cdot x_i$.

## 6. Conclusion

We have presented a scalable, energy-efficient optical neural network with customized multi-operand optical neurons (MOONs). We have experimentally demonstrated a 4-operand silicon-photonic MOMZI on practical image recognition tasks. Compared to prior single-operand-MZI-based photonic tensor cores (PTCs), our MOMZI-based PTC design can achieve one to two orders-of-magnitude reduction in active optical component usage, footprint, latency and propagation loss. The speed, footprint, and energy efficiency of our MOON-based PTC can benefit from more advanced technologies, e.g., faster and more efficient data converters, optical devices, and nonlinearity engineering. Our customized MOON design provides a scalable solution for the next-generation photonic AI accelerators with extreme compute density and energy efficiency.

## 7. Supporting information

### 7.1 Parameter tables for the delay, propagation loss, and footprint estimation

**Table. S1.** Device parameters used in our performance estimation based on AIM photonics' PDK [15].

| Optical component | Length (μm) | Insertion loss (dB) |
|---|---|---|
| High-speed EO MZI ($MZI(hs)$) | 1600 | 3 |
| High-speed plasmonic EO MZI [19] | ~220[a] | 11.2 |
| Low-speed TO MZI ($MZI(ls)$) | 550 | 1 |
| Microring-based filter | 16 | 0.25 |

a. Based on the layout picture of the MZI in the reference (~$200 \times 100$ $\mu m^2$). The high-speed phase shifter part is only 15 $\mu m$ in length, so the size of the modulator can be further optimized with more compact directional couplers.

**Table. S2.** Parameters to calculate the performance of $k$-op MOMZI-PTC.

| Parameters | Values |
|---|---|
| $L_{MOMZI}$ | $L_{MZI(hs)} + (k-1)d$ |
| $L_{combiner}$ | $\frac{n}{k}L_{ring}$ |
| $d$ | 10 $\mu m$ (1.5 $\mu m$ after scaling) |
| $IL_{MOON}$ | $IL_{MZI(hs)}$ |
| $IL_{combiner}$ | $\frac{n}{k}IL_{ring}$ |
| $S_{MOMZI}$ | $L_{MOMZI} \times W_{MZI}(W_{MOMZI} = 460 \ \mu m)$ |
| $S_{combiner}$ | $\frac{mn}{k} \times L_{ring} \times W_{ring}(W_{ring} = 16 \ \mu m)$ |
| $S_{MZI(hs)}$ | $L_{MZI(hs)} \times W_{MZI(hs)} \ (MZI(hs) = 460 \ \mu m)$ |
| $S_{MZI(ls)}$ | $L_{MZI(ls)} \times W_{MZI(ls)} \ (W_{MZI(ls)} = 127 \ \mu m)$ |

## 7.2 Energy efficiency.

The power consumption of $n$-input, $m$-output MOMZI-PTC for computing is contributed by lasers, weight configuration, and conversion between electrons and photons, which is obtained by:

$$P_{MOMZI-ONN} = P_{laser} + n(\frac{m}{k}E_{MOMZI} + E_{DAC})f_X + \frac{mn}{k}P_{thermal} + mP_{ADC} \quad \text{Eq. (S1)}$$

The parameter table for modeling Eq. (S1) is provided in Table. S3. In Eq. (S1), $P_{laser}$ is the laser power. $E_{MOMZI}$ represents the energy consumption of modulators, $E_{DAC}$ and $E_{oe}$ include the power consumption for photodetection, amplification, and analog-to-digital conversion. $f_{md}$ is the operating speed of modulators, as determined by the total delay of the MOMZI-ONN. In the MOMZI-ONN, $P_{thermal}$ is the static power to tune the MOMZI to a bias point. Using thermal phase shifters for bias tuning, $P_{thermal} = 2.5 \ mW$. By utilizing energy-efficient active optical components based on nano-opto-electro-mechanical systems or phase change materials [46,47], we can eliminate the power consumption for phase maintenance. To carrier-depletion-based silicon MZI modulators, $E_{MOMZI}$ can achieve ~146 fJ/bit. Using energy-efficient plasmonic-on-silicon modulators, $E_{MOMZI} = 0.1$ fJ/bit. One DAC's energy consumption can be estimated by [48]:

$$E_{DAC} = F_D n_b F_s / B_r \quad \text{Eq. (S2)}$$

where $F_D$ is the DAC figure of merit, $n_b$ is the DAC resolution, $F_s$ is the sampling frequency, $B_r$ is the bit rate. In our estimation, $F_D = 35$ fJ/step in a 7-nm microprocessor [42], $n_b = 8$ bit, $F_s/B_r = 1$.

The propagation loss, the photodetectors' minimum detectable power, and the outputs' precision dictate the laser power. The total laser power can then be calculated by the following equation [6]:

$$P_{laser} = m\left(\frac{n}{\rho^2}\frac{h\nu}{\eta IL}\max\left(2^{2N_b+1}, \frac{C_d V_r}{e}\right)f_{md}\right) \times \frac{n}{k} \quad \text{(Eq. S3)}$$

where $h\nu$ is the photon energy at 1.55 μm, $\rho = n, \eta = 0.2$ is the wall-plug efficiency of the laser [49], $\frac{n}{k}$ is the number of wavelengths used in the MOMZI-PTC. The precision of output signals is $N_b$ bits. $C_d$ is the capacitance of the photodetector while $V_r$ is the operating voltage. Note that $V_r = 0$ in some zero-biased energy-efficient photodetectors [50,51], $f_X$ is the baud rate of the intput signals. In this work, we choose $f_X = 10G$ Baud/s and use a 10 GSPS ADC for reading the output.

**Table. S3.** Energy consumption of a $k$-point $m \times n$ MOMZI-PTC – modeling parameters

| Expression | Value |
|---|---|
| $E_{MOMZI}$ | ~146 fJ/bit [31] |
|  | (0.1 fJ/bit after scaling) [19] |
| $P_w$ | 2.5 $mW$ (PDK) |
|  | 0 after scaling |
| $E_{DAC} = F_D N_b F_s/B_r$ [48] | $F_D = 35$ fJ/step [42] |
|  | $N_b = 8\ bit$ |
|  | $F_s/B_r = 1$ |
| $P_{laser} = \frac{mn}{k}\left(\frac{n}{\rho^2}\frac{h\nu}{\eta IL}\max\left(2^{2N_b+1}, \frac{C_d V_r}{e}\right)f_{md}\right)$ [6] | $\rho = n$ |
|  | $\nu = 193.5\ THz$ |
|  | $\eta = 0.2$ |
|  | $V_r = 0$ |
| $P_{ADC}$ (10 GSPS) | 39 mW/channel [52] |
|  | 0.52 fJ/level [37] (1.3 mW/channel) after scaling |

*Acknowledgments*

The authors acknowledge support from the Multidisciplinary University Research Initiative (MURI) program through the Air Force Office of Scientific Research (AFOSR), monitored by Dr. Gernot S. Pomrenke.

**Competing interests**

The authors declare no competing interests.

**Data availability**

The data and codes that support the findings of this study are available from the corresponding author upon reasonable request.